\documentclass[]{spie}  

\usepackage{amsmath,amsfonts,amssymb}
\usepackage{graphicx}
\usepackage{multirow}
\usepackage{caption}
\usepackage{subcaption}
\usepackage[colorlinks=true, allcolors=blue]{hyperref}
\usepackage{orcidlink}

\title{The Simons Observatory: Studies of Detector Yield and Readout Noise From the First Large-Scale Deployment of Microwave Multiplexing at the Large Aperture Telescope}

\author[a,b]{Thomas P. Satterthwaite\,\orcidlink{0000-0002-6452-4220}}
\author[b,c]{Zeeshan Ahmed\,\orcidlink{0000-0002-9957-448X}}
\author[d]{Kyuyoung Bae\,\orcidlink{0000-0002-3376-8660}}
\author[e]{Mark Devlin\,\orcidlink{0000-0002-3169-9761}}
\author[e]{Simon Dicker\,\orcidlink{0000-0002-1940-4289}}
\author[f]{Shannon M. Duff\,\orcidlink{0000-0002-9693-4478}}
\author[g]{Daniel Dutcher\,\orcidlink{0000-0002-9962-2058}}
\author[e]{Saianeesh K. Haridas\,\orcidlink{0000-0001-6519-502X}}
\author[b,c]{Shawn W. Henderson\,\orcidlink{0000-0001-7878-4229}}
\author[f]{Johannes Hubmayr}
\author[h]{Bradley R. Johnson\,\orcidlink{0000-0002-6898-8938}}
\author[e]{Anna Kofman\,\orcidlink{0000-0001-5374-1767}}
\author[i]{Jack Lashner\,\orcidlink{0000-0002-6522-6284}}
\author[f]{Michael J. Link\,\orcidlink{0000-0003-2381-1378}}
\author[f]{Tammy J. Lucas\,\orcidlink{0000-0001-7694-1999}}
\author[e]{Alex Manduca\,\orcidlink{0000-0003-4629-5759}}
\author[j,k]{Michael D. Niemack\,\orcidlink{0000-0001-7125-3580}}
\author[e]{John Orlowski-Scherer\,\orcidlink{0000-0003-1842-8104}}
\author[b,c]{Tristan Pinsonneault-Marotte\,\orcidlink{0000-0002-9516-3245}}
\author[i]{Max Silva-Feaver}
\author[g]{Suzanne Staggs\,\orcidlink{0000-0002-7020-7301}}
\author[j,l]{Eve M. Vavagiakis}
\author[g]{Yuhan Wang\,\orcidlink{0000-0002-8710-0914}}
\author[g]{Kaiwen Zheng\,\orcidlink{0000-0003-4645-7084}}
\affil[a]{Department of Physics, Stanford University; Stanford, CA 94305; USA}
\affil[b]{Kavli Institute for Particle Astrophysics and Cosmology; Stanford, CA 94305; USA}
\affil[c]{SLAC National Accelerator Laboratory; Menlo Park, CA 94025; USA}
\affil[d]{University of Colorado Boulder; Boulder, CO, 80309; USA}
\affil[e]{Department of Physics and Astronomy, University of Pennsylvania; Philadelphia, PA 19104; USA}
\affil[f]{Quantum Sensors Division, National Institute of Standards and Technology; Boulder, CO, 80305; USA}
\affil[g]{Joseph Henry Laboratories of Physics, Princeton University; Princeton, NJ 08542; USA}
\affil[h]{Department of Astronomy, University of Virginia; Charlottesville, VA 22904; USA}
\affil[i]{Wright Laboratory, Department of Physics, Yale University; New Haven, CT 06520; USA}
\affil[j]{Department of Physics, Cornell University; Ithaca, NY 14853; USA}
\affil[k]{Department of Astronomy, Cornell University; Ithaca, NY 14853; USA}
\affil[l]{Department of Physics, Duke University; Durham, NC 27708; USA}

\authorinfo{Corresponding author: Thomas P. Satterthwaite\\Email: tpsatt@stanford.edu}

\begin{document} 
\maketitle

\begin{abstract}
The Simons Observatory is a new ground-based cosmic microwave background experiment, which is currently being commissioned in Chile’s Atacama Desert. During its survey, the observatory’s small aperture telescopes will map 10\% of the sky in bands centered at frequencies ranging from 27 to 280\,GHz to constrain cosmic inflation models, and its large aperture telescope will map 40\% of the sky in the same bands to constrain cosmological parameters and use weak lensing to study large-scale structure. To achieve these science goals, the Simons Observatory is deploying these telescopes' receivers with 60,000 state-of-the-art superconducting transition-edge sensor bolometers for its first five year survey. Reading out this unprecedented number of cryogenic sensors, however, required the development of a novel readout system. The SMuRF electronics were developed to enable high-density readout of superconducting sensors using cryogenic microwave SQUID multiplexing technology. The commissioning of the SMuRF systems at the Simons Observatory is the largest deployment to date of microwave multiplexing technology for transition-edge sensors. In this paper, we show that a significant fraction of the systems deployed so far to the Simons Observatory's large aperture telescope meet baseline specifications for detector yield and readout noise in this early phase of commissioning.
\end{abstract}


\section{INTRODUCTION}
\label{sec:intro}  

The Simons Observatory is a cosmic microwave background (CMB) experiment which is currently being commissioned at an elevation of 5,200\,m on Cerro Toco in Chile's Atacama Desert. The experiment presently consists of three 0.5\,m refracting small aperture telescopes (SATs) and a single 6\,m Crossed Dragone large aperture telescope (LAT), with bandpasses ranging from 27 to 280\,GHz.\cite{SO-Forecasts} The SATs will map 10\% of the sky targeting 2\,$\mu$K-arcmin map noise in 90 and 150\,GHz bands with the aim of searching for $B$-mode polarization to constrain cosmic inflation models. The LAT will map 40\% of the sky targeting 6\,$\mu$K-arcmin map noise in 90 and 150\,GHz bands with the aims of constraining cosmological parameters, studying the Universe's large-scale structure via weak gravitational lensing, and constraining the sum of the masses of neutrinos. The Simons Observatory collaboration has constructed all four telescopes on Cerro Toco, and the LAT is expected to achieve first light in early 2025.

\begin{figure}[t]
    \centering
    \includegraphics[width=0.66\textwidth]{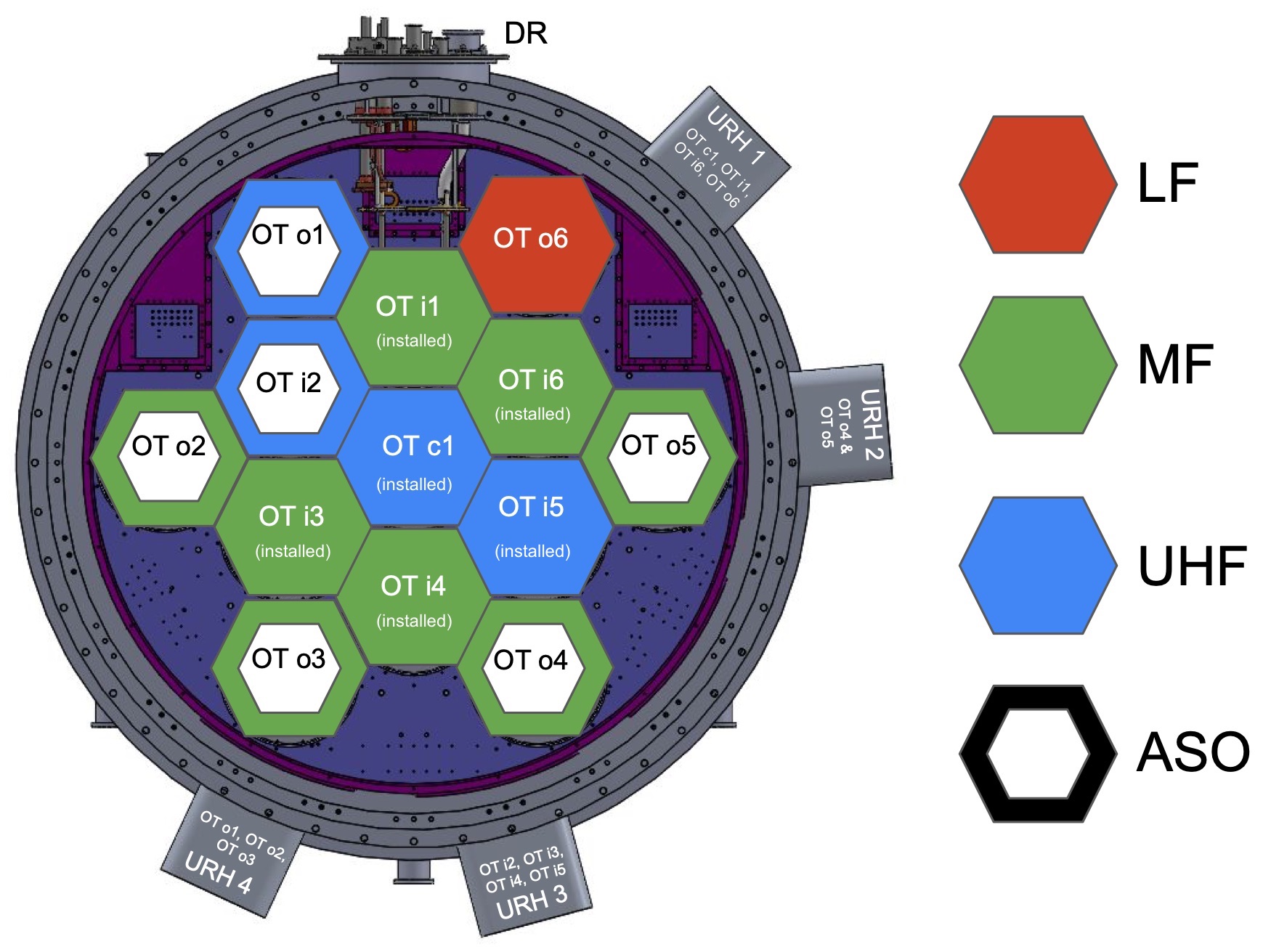}
    \caption{Rear-view schematic showing the positions of the LAT receiver's optics tubes (OTs), with low-frequency (LF), mid-frequency (MF), and ultra-high-frequency (UHF) OTs labelled. The LF OT and those slated for the Advanced Simons Observatory (ASO) upgrade have not yet been installed. Positions of the dilution refrigerator (DR), which provides cooling, and the universal readout harnesses (URHs), which support the DC lines and RF coaxial channels, are also shown. Credit: Bob Thornton/University of Pennsylvania LATR team.}
    \label{fig:latr}
\end{figure}

Constructing the Simons Observatory requires fabricating and deploying 60,000 state-of-the-art superconducting transition-edge sensor (TES) bolometers. These TESs are split across the four telescopes with roughly 10,000 in each SAT and 30,000 in the LAT.\cite{SATs}\cite{LATR} The detectors are deployed in universal focal-plane modules (UFMs), each of which contains up to 1,756 TESs (of which up to 1,720 are optically coupled) and 28 multiplexer chips.\cite{UFM} These multiplexer chips use microwave superconducting quantum interference device (SQUID) multiplexing ($\mu$mux) in order to achieve multiplexing factors reaching 910 and enable high-density readout; this is a dramatic improvement over previous iterations of time- and frequency-division multiplexing technologies which have each been able to reach a maximum factor of 68 when deployed for CMB science.\cite{MatesThesis}\cite{TDM}\cite{SPT-3G} This has been enabled by the large-scale fabrication of TESs and microwave SQUIDs, and by the development of the SLAC Microresonator RF (SMuRF) electronics, a novel signal transduction and readout system.\cite{SO-Wafers}\cite{SO-MuxChips}\cite{SMuRF}

In $\mu$mux, each TES is inductively coupled to a SQUID which is in turn coupled to a microwave frequency resonator such that the change in TES current induced by observed signals perturbs the resonator's frequency. The SMuRF system provides the integrated electronics for serving the tones which drive and track the resonators, the flux ramp signal which linearizes the SQUID response, and the DC signal which biases the TESs. This therefore enables the deployment, control, and data acquisition of $\mu$mux technology at scale.

The construction of the Simons Observatory has been the largest deployment to date of a $\mu$mux system for astronomy; previous deployments of $\mu$mux at MUSTANG-2 and the Keck Array, for example, have read out 215 and 528 detectors, respectively.\cite{MUSTANG223}\cite{BKumux} As 7 UFMs have been deployed to each SAT and 18 have been deployed to the LAT, a commensurate number of SMuRF systems, with each one able to provide the warm readout electronics for a single UFM, have also been deployed. As such, an important part of ensuring the readiness of these systems and their UFMs for CMB observation is measuring the detector channel yield and readout noise, and ensuring that these statistics match required performance specifications. The following section of this paper discusses ongoing work towards this end for the Simons Observatory's LAT.

\section{Readout Commissioning}
\label{sec:readout}

\begin{figure}
    \centering
    \begin{subfigure}[b]{0.4\textwidth}
        \centering
        \includegraphics[width=\textwidth]{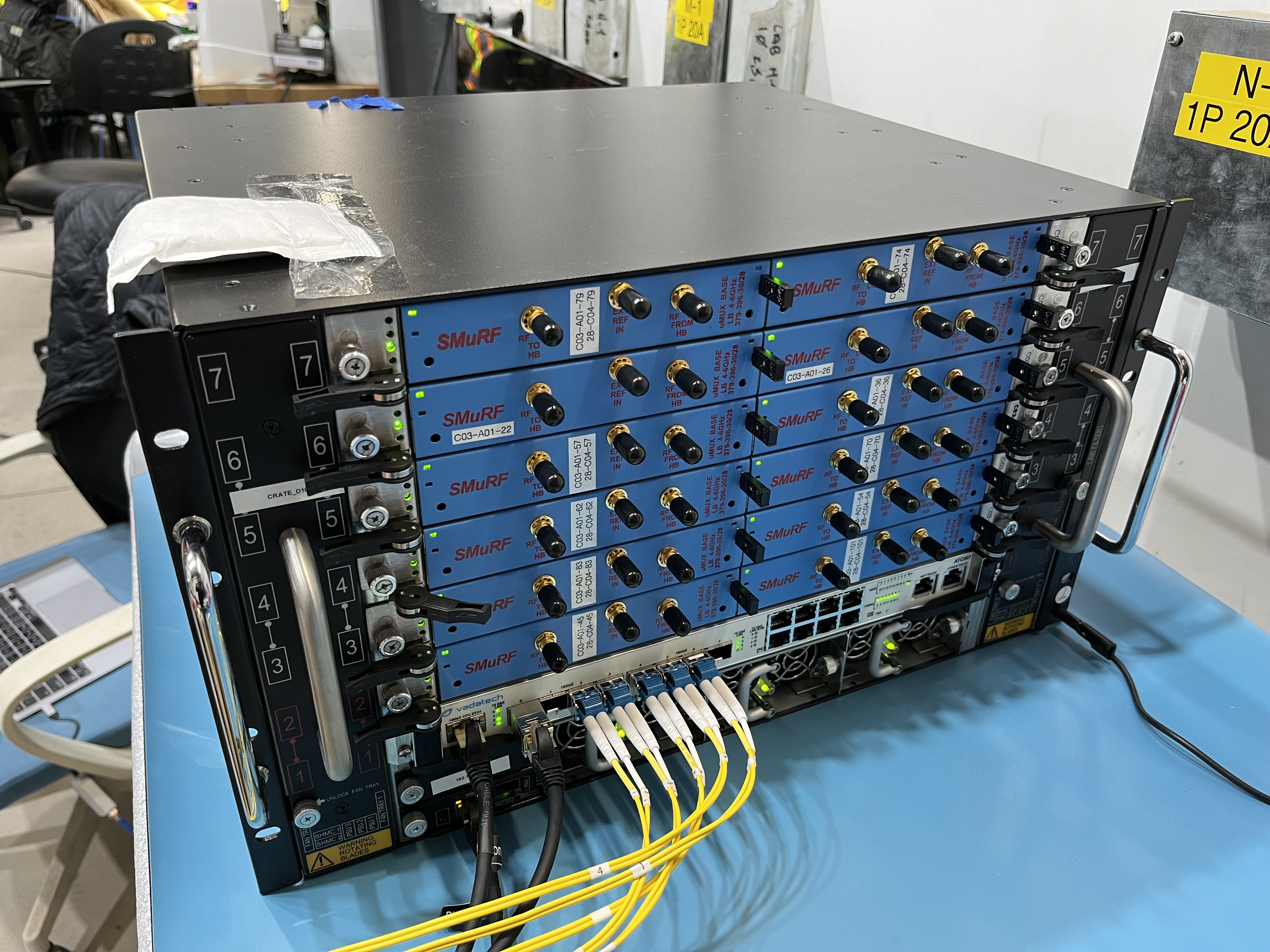}
        \caption{}
        \label{fig:smurf-bench}
    \end{subfigure}
    \begin{subfigure}[b]{0.4\textwidth}
        \centering
        \includegraphics[width=\textwidth]{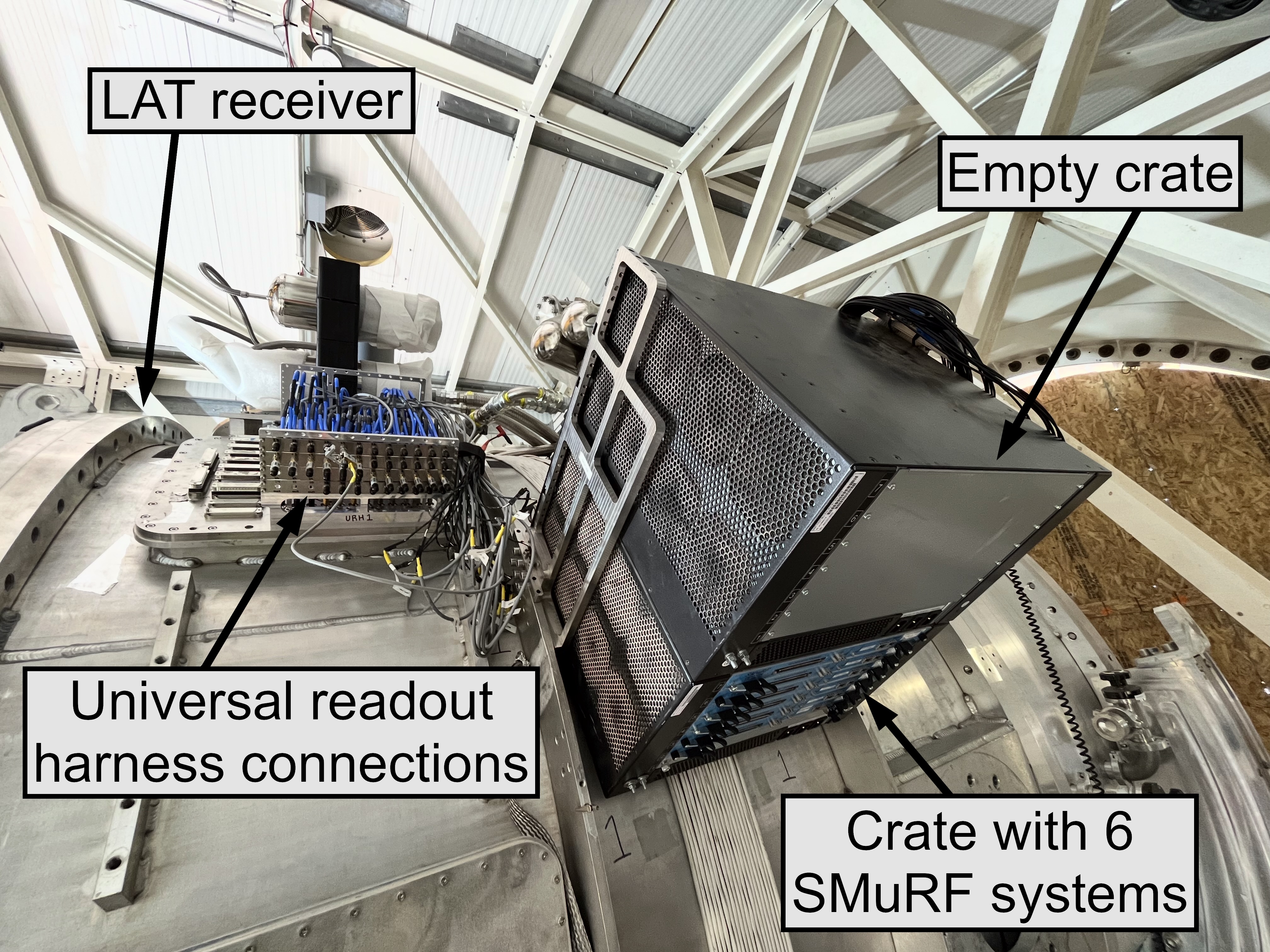}
        \caption{}
        \label{fig:smurf-latr}
    \end{subfigure}
    \caption{(a) One crate containing six SMuRF systems on a test bench in the Simons Observatory high bay lab facility on Cerro Toco, Chile. (b) Two crates, one populated with six SMuRF systems and the other empty, mounted to the LATR (without fiber, RF coaxial, and DC cables installed).}
    \label{fig:smurf}
\end{figure}

The deployment of the LAT's optics tubes (OTs) to the cryogenic LAT receiver (LATR) has occurred in phases. In mid-2023, two OTs (identified as OTi1 and OTi6) were installed and capped at 4\,K. In early 2024, four more OTs (identified as OTc1, OTi3, OTi4, and OTi5) were installed, and all six OTs were capped at 300\,K. The positions of these OTs are shown in Figure \ref{fig:latr}. Each OT contains three UFMs for a current total of 18 UFMs containing more than 30,000 TESs.\cite{LATR} Along with each deployment of UFMs, a commensurate number of SMuRF systems have also been installed in the LAT to provide the warm readout electronics for the detector modules. One crate containing six SMuRF systems is shown in Figure \ref{fig:smurf-bench}. As of early 2024, four such crates, each containing between three and six SMuRF systems, have been mounted to the exterior of the LATR, as shown in Figure \ref{fig:smurf-latr}, so that they are able to corotate with the receiver as the elevation structure rotates. The next planned personnel deployment to the LAT site is scheduled for mid-2024, when we will continue the commissioning tasks which require in-person operations.

As the LAT's mirrors are expected to be delivered in early 2025, each optics tube is currently capped at 300\,K with an aluminum plate. Nonetheless, characterization of the dark performance of the UFMs and SMuRF systems during this phase of commissioning provides valuable information and statistics about detector and readout performance. As we carry out these activities, we seek to meet baseline specifications for the Simons Observatory's operation, including a detector yield in excess of 70\%, with the goal of reaching 80\%, and readout noise equivalent current below the level of 65\,pA/$\sqrt{\text{Hz}}$, with the goal of reaching 45\,pA/$\sqrt{\text{Hz}}$. Meeting these specifications will enable us to meet the Simons Observatory's science goals.\cite{SO-Forecasts}

Since the departure of the LATR instrument team in early 2024, we have seen a few fixable hardware-related issues with the telescope's readout system. During the re-installation of OT OTi6 in early 2024, there was an accidental swap involving its cold coaxial cabling, so this OT's UFMs cannot currently be read out. This OT was, however, operational in mid-2023 when it was capped at 4\,K, so we are able to present results from that phase of commissioning. Since early 2024, we have also seen a few SMuRF hardware-related issues. On the SMuRF system reading out UFM Mv28, we find that the RF tone power being applied is inadequate due to an RF tone generation hardware problem. We plan to replace the relevant SMuRF part in mid-2024. On the SMuRF systems reading out UFMs Mv21 and Mv24, the SMuRF DC cables appear to be inadequately engaged. We have seen and fixed this issue in the past, so we plan on correcting it during our next trip to the site by reseating and securing, or possibly replacing, the connectors. Because UFMs Mv21, Mv24, and Mv28 were all deployed to the LATR in mid-2023, we are still able to present results from their initial commissioning, when their OTs were capped at 4\,K. Finally, we have also seen hardware issues relating to the DC and low-frequency biasing systems of the SMuRF systems which read out UFMs Mv34 and Mv49. We plan to correct these issues by replacing the relevant SMuRF parts in mid-2024, however we are not yet able to present any commissioning data from these systems.

\subsection{Detector Yield}
\label{sec:yield}

We measure detector yield by determining the number of detectors which exhibit visible, smooth transitions from normal to superconducting states. We determine this total yield by using SMuRF to locate and track the resonators to which the detectors are coupled, and to measure the detectors' responses to input electrical power. Note that, as defined, this total yield is the net effect of detector-only and readout-only losses.

This process first requires determining the correct power at which to send the probe tones which interrogate each resonator. To do this, we can adjust the tone power itself, as well as the attenuations of the upconverters and downconverters which come before and after the signal enters and returns from the cryostat, respectively. Leaving the tone power fixed, as it strongly influences the readout noise which is discussed in the following section, we have determined these attenuations by finding the combination which minimizes the streamed noise level from the detectors. With the probe tone power properly optimized, shifts in the frequencies of these resonators, as would be induced by detector signals, can then be tracked via SMuRF's closed-loop tracking algorithm.

\begin{figure}
    \centering
    \begin{subfigure}[b]{0.45\textwidth}
        \centering
        \includegraphics[width=\textwidth]{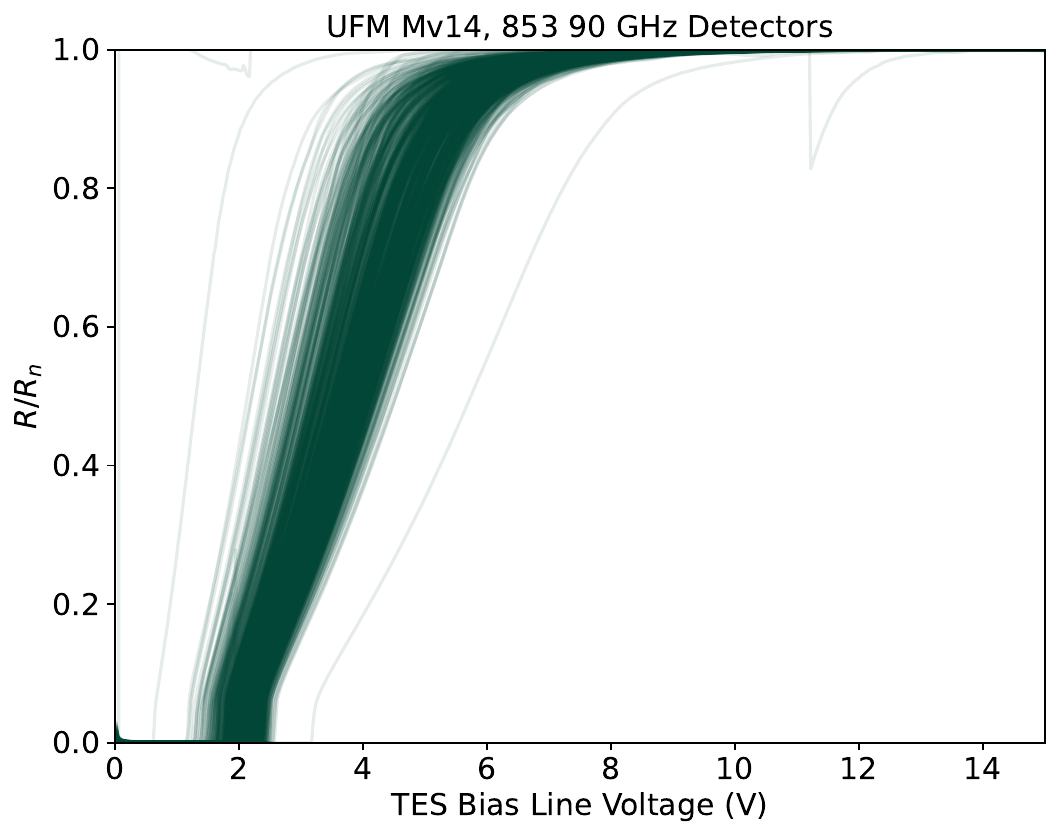}
        \caption{}
        \label{fig:ivs-90}
    \end{subfigure}
    \begin{subfigure}[b]{0.45\textwidth}
        \centering
        \includegraphics[width=\textwidth]{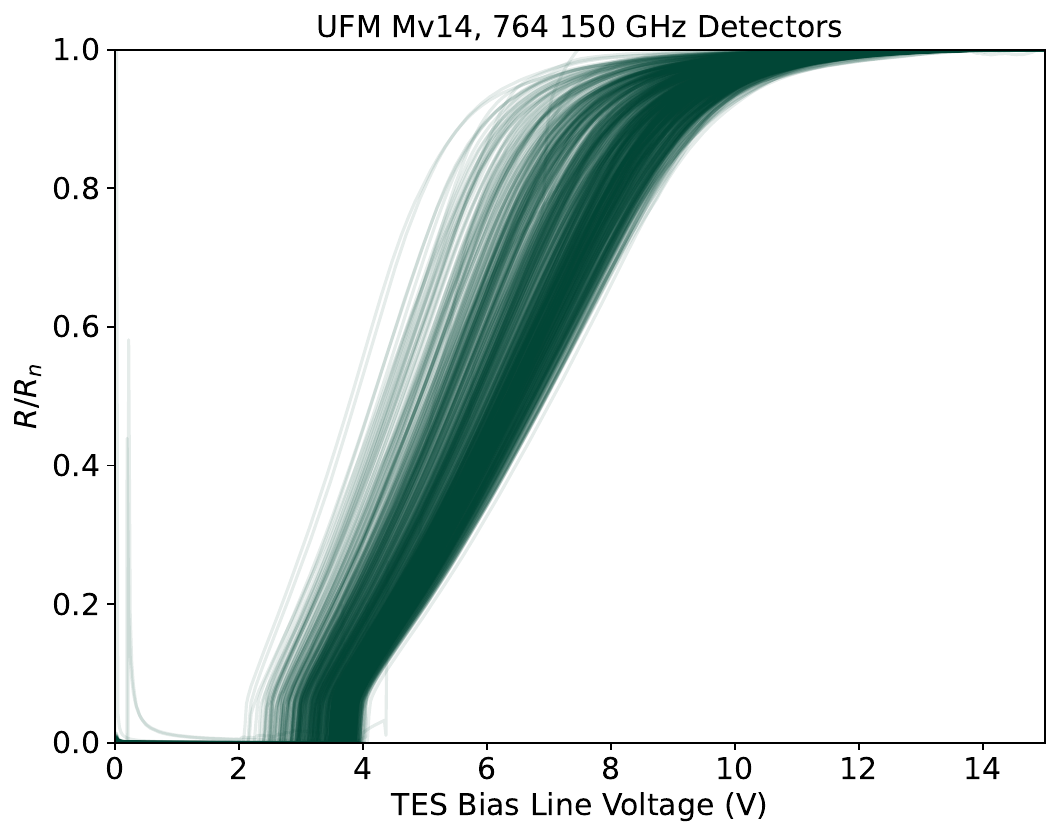}
        \caption{}
        \label{fig:ivs-150}
    \end{subfigure}
    \caption{Curves showing TES resistance (relative to normal resistance) versus bias line voltage for the (a) 90\,GHz and (b) 150\,GHz detectors of a sample UFM (Mv14).}
    \label{fig:ivs}
\end{figure}

Once resonances have been located and tracked, we characterize the detectors by taking $I{-}V$ curves. These measure the response of each TES (as read out through the SQUIDs to which they are coupled) as a range of voltages are applied to their bias lines, which have an average resistance of about 16\,k$\Omega$, and through the TESs, which have a normal resistance ($R_n$) of about 7 to 9\,m$\Omega$ and are shunted by resistors of approximately 400\,$\mu\Omega$. This creates curves as shown in Figure \ref{fig:ivs} (scaled to $R/R_n$), showing the transition between normal and superconducting behavior. A detector is then considered part of the yield if it demonstrates a visible, smooth transition from a flat normal region to a superconducting state as determined by basic quality cuts. We have taken these curves for 16 UFMs deployed to the LAT, and results are summarized in Table \ref{tab:ufm-yield}.

Due to the hardware issues discussed in Section \ref{sec:readout} which affect two SMuRF DC and low-frequency biasing systems, results are not presented from their associated UFMs (identified as Mv34 and Mv49). Due to the cold coaxial cabling, RF tone generation, and DC cable engagement issues discussed in the same section, results presented for UFMs Mv11, Mv21, Mv24, Mv25, Mv26, and Mv28 are from studies performed prior to the early 2024 deployment, when their OTs were capped at 4\,K. The procedure was, however, the same as that used for the ten UFMs which were surveyed later in 2024, when their OTs were capped at 300\,K. We expect that the change in optical loading has a limited effect on the detector yield as determined using this method, so the yield would therefore be consistent for 300\,K caps.

\begin{table}[t]
\caption{Detector yield from 16 UFM and SMuRF systems.} 
\label{tab:ufm-yield}
\begin{center}       
\begin{tabular}{|l|l|l|l|} 
\hline
\rule[-1ex]{0pt}{3.5ex}  Optics Tube Identifier & UFM Identifier & Detector Yield (\#) & Detector Yield (\%)  \\
\hline
\rule[-1ex]{0pt}{3.5ex}  \multirow{3}{*}{OTi1} & Mv28 & 1496 & 85.2 \\\cline{2-4}

\rule[-1ex]{0pt}{3.5ex}                        & Mv24 & 1516 & 86.3 \\\cline{2-4}

\rule[-1ex]{0pt}{3.5ex}                        & Mv21 & 1423 & 81.0 \\
\hline
\rule[-1ex]{0pt}{3.5ex}  \multirow{3}{*}{OTi6} & Mv25 & 1629 & 92.8 \\\cline{2-4}

\rule[-1ex]{0pt}{3.5ex}                        & Mv11 & 1306 & 74.4\\\cline{2-4}

\rule[-1ex]{0pt}{3.5ex}                        & Mv26 & 1493 & 85.0 \\
\hline
\rule[-1ex]{0pt}{3.5ex}  \multirow{2}{*}{OTi3} & Mv13 & 1057 & 60.2 \\\cline{2-4}

\rule[-1ex]{0pt}{3.5ex}                        & Mv20 & 1589 & 90.5 \\
\hline
\rule[-1ex]{0pt}{3.5ex}  \multirow{2}{*}{OTi4} & Mv14 & 1617 & 92.1 \\\cline{2-4}

\rule[-1ex]{0pt}{3.5ex}                        & Mv32 & 1555 & 88.6 \\
\hline
\rule[-1ex]{0pt}{3.5ex}  \multirow{3}{*}{OTi5} & Uv42 & 1503 & 85.6 \\\cline{2-4}

\rule[-1ex]{0pt}{3.5ex}                        & Uv31 & 1504 & 85.6 \\\cline{2-4}

\rule[-1ex]{0pt}{3.5ex}                        & Uv47 & 1557 & 88.7 \\
\hline
\rule[-1ex]{0pt}{3.5ex}  \multirow{3}{*}{OTc1} & Uv46 & 1437 & 81.8 \\\cline{2-4}

\rule[-1ex]{0pt}{3.5ex}                        & Uv39 & 1542 & 87.8 \\\cline{2-4}

\rule[-1ex]{0pt}{3.5ex}                        & Uv38 & 1300 & 74.0 \\
\hline
\end{tabular}
\end{center}
\end{table}

We see that the detector yield of 15 of the presented UFM and SMuRF systems is in excess of 70\% of deployed TESs. We find these results promising as they satisfy the baseline performance requirement, and 13 UFMs satisfy the goal requirement of 80\% yield. Furthermore, 14 UFMs show detector yields which are within 10\% of pre-deployment screening numbers, with potential gains available when the remaining detector modules are assessed and further optimized.\cite{UFM_MF}\cite{UFM_LF} Work is ongoing to understand and potentially correct the insufficient yields of the UFMs which do not meet the goal specification. UFM Mv11's yield is known to be low because of a problem with a TES bias line on the module, and we plan to repair this issue. For other UFMs, insufficient yield could be due to missing or collided resonators, to detectors which are not tracking well due to SQUID properties, or to damaged wire bonds or traces.

\subsection{Readout Noise}
\label{sec:noise}

Given that the LAT's TESs are not currently optically loaded by the sky, when they are normal (off-transition), noise is dominated by the quadrature sum of their Johnson noise and the readout noise.\cite{UFM} We can therefore infer the readout noise equivalent current by subtracting the known contribution of Johnson noise from the median spectral white noise level that we compute between 10 and 30\,Hz. With normal TESs, the low-frequency limit of the Johnson noise current spectral density is given by:

\[\text{NEI}_\text{Johnson}=\sqrt{\frac{4k_\text{B}\left(T_{S}R_{S}+T_\text{TES}R_\text{TES}\right)}{\left(R_{S}+R_\text{TES}\right)^2}}\]

Where $T_S$ is the temperature of the TES's shunt resistor (which we approximate to be the bath temperature, $T_\text{Bath}$), $R_{S}$ is the resistance of said shunt resistor, $T_\text{TES}$ is the TES's temperature, $R_\text{TES}$ is the TES's normal resistance, and $k_B$ is the Boltzmann constant.

\begin{figure}
    \centering
    \includegraphics[width=0.5\textwidth]{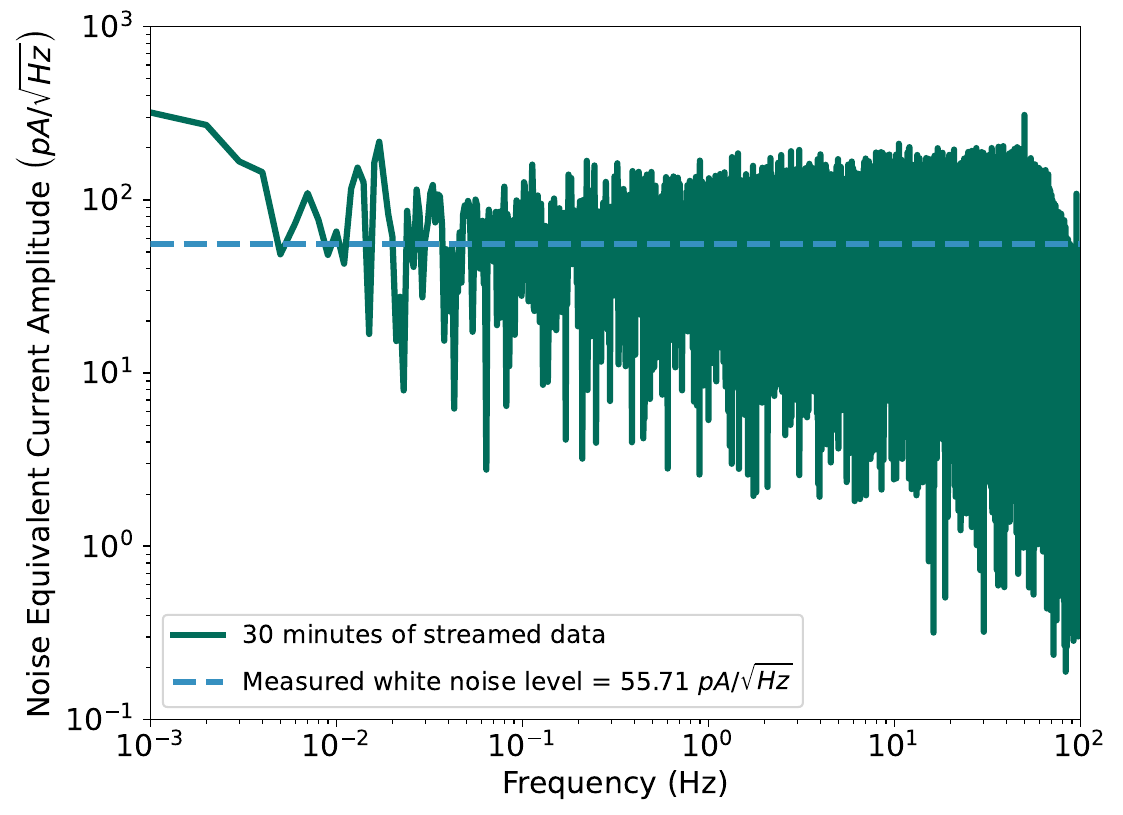}
    \caption{Amplitude spectral density of 30 minutes of streamed data from a sample detector whose white noise level is close to the overall median. Peak at 50\,Hz is due to the frequency of Chilean AC power. Amplitude spectral density is cut off at 100\,Hz due to the Nyquist sampling of the readout filter which is applied. This also causes the roll-off towards higher frequencies.}
    \label{fig:tod}
\end{figure}

For OTs OTc1, OTi3, OTi4, and OTi5, we inferred readout noise by recording 30 minute-long streams sampled at 200\,Hz with the detectors normal. These detector conditions were achieved by setting the bath temperature higher than the TESs' critical temperatures (typically $T_c\approx165$\,mK and $T_\text{Bath}\approx220$\,mK).\cite{UFM_MF}\cite{UFM_LF} A sample amplitude spectral density of data from one detector taken using this method is shown in Figure \ref{fig:tod}. We then calculated and subtracted off the Johnson noise using this bath temperature for $T_S$ and $T_\text{TES}$, and the per-UFM median $R_\text{TES}$ as measured prior to deployment to Chile.\cite{UFM_MF}\cite{UFM_LF}

\begin{figure}
    \centering
    \includegraphics[width=0.80\textwidth]{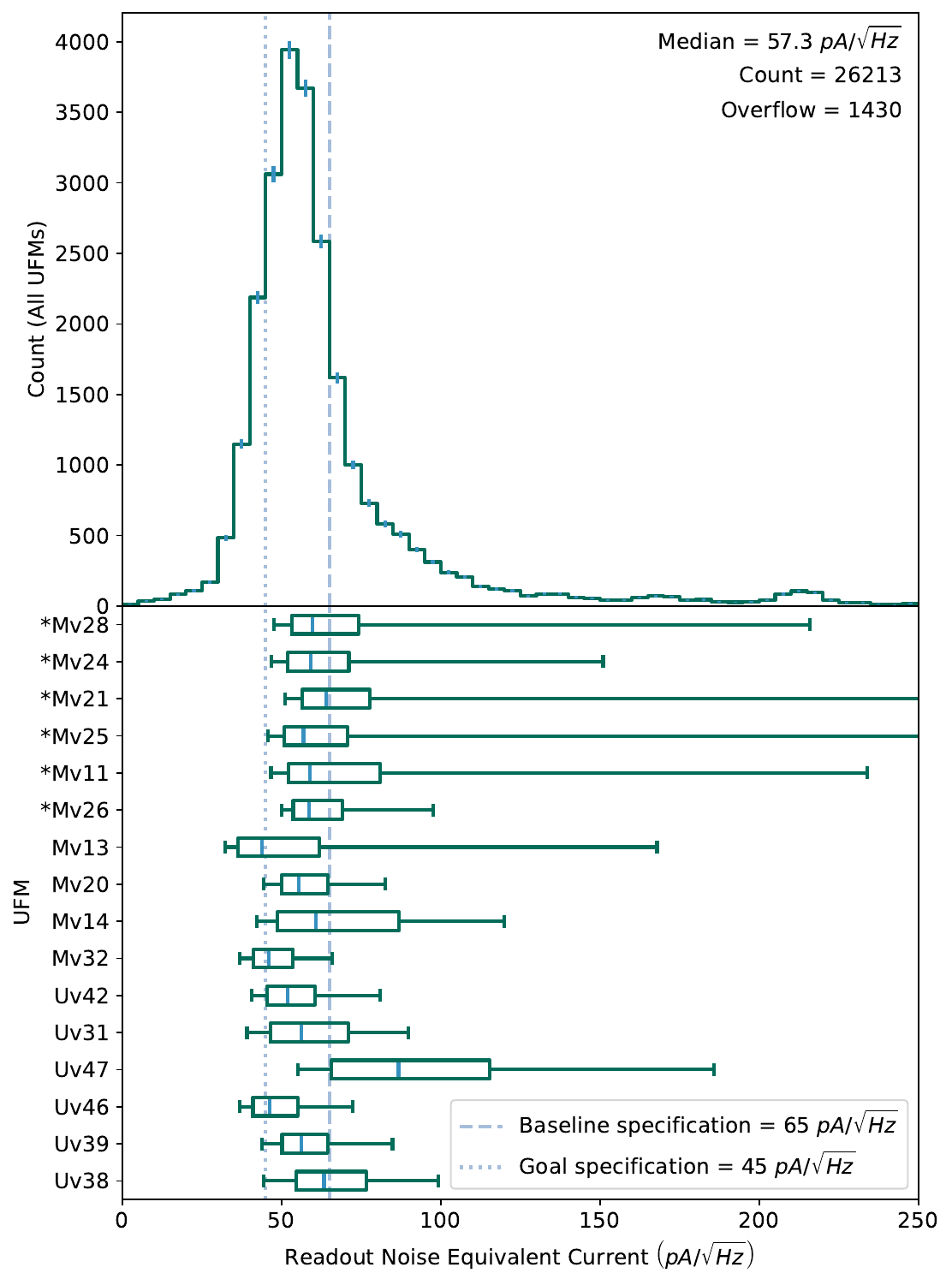}
    \caption{Readout noise equivalent current calculated from time streams with the detectors normal for 16 UFMs. Note that for the UFMs marked with asterisks, the detectors were made normal by driving saturation current down their bias lines, while for the other UFMs, the detectors were made normal by raising the bath temperature above their critical temperatures. Top panel shows the distribution of all detectors on the 16 UFMs, while the bottom panel shows the distribution within each UFM. Boxes show the second and third quartiles of data with medians marked by vertical blue lines, while the whiskers bound the central 80\% of data. Vertical dashed and dotted lines show the baseline and goal specifications, respectively.}

    \label{fig:readout-noise}
\end{figure}

Due to the cold coaxial cabling, RF tone generation, and DC cable engagement issues discussed in Section \ref{sec:readout}, we were not able to use this procedure for OTs OTi1 and OTi6 which contain UFMs Mv11, Mv21, Mv24, Mv25, Mv26, and Mv28. Instead, we present results from late 2023 when the systems were operational. In this test, the detectors were made normal by driving saturation current through each of the UFMs' TES bias lines in order to induce sufficient heating. This brought $T_\text{TES}$ above the detectors' critical temperatures, which were measured prior to deployment to Chile. We could then approximate and subtract off the Johnson noise using the per-UFM median values of these critical temperatures and of $R_\text{TES}$, which was also measured prior to deployment. We intend to re-characterize the readout noise of these UFMs using longer streams and a raised bath temperature after the hardware issues are addressed this summer.

Results for the calculated readout noise from 16 UFMs are shown in Figure \ref{fig:readout-noise}. Due to the SMuRF DC and low-frequency biasing system issues discussed in Section \ref{sec:readout}, results are not presented from UFMs Mv34 and Mv49. For the other UFMs, we see that there is not a significant difference between the calculated readout noise levels as given by driving saturation current through the TESs' bias lines (as denoted by asterisks in the figure) and that given by raising the bath temperature. We find that 15 of the surveyed UFMs show median readout noise levels below 65\,pA/$\sqrt{\text{Hz}}$. This is a promising result as it shows performance consistent with the baseline specification for science operation. For UFM Uv47, which shows an anomalously high readout noise level, we are investigating this issue.

\section{Conclusion}
\label{sec:conclusion}

These results show promising progress towards commissioning the detectors and readout systems of the Simons Observatory's LAT. While some of the deployed UFMs and SMuRF systems are affected by understood readout hardware issues, a large portion of the remaining systems show detector yield and readout noise levels which are consistent with the observatory's baseline specifications for science operation. As we bring the remaining systems online later in 2024, we will employ similar procedures to characterize these systems and ensure their readiness for operation.

\acknowledgments 
 
This work was supported in part by the Department of Energy, Laboratory Directed Research and Development program at SLAC National Accelerator Laboratory, under contract DE-AC02-76SF00515. This work was also supported in part by a grant from the Simons Foundation (Award \#457687, B.K.). This work was also supported by the National Science Foundation (UEI GM1XX56LEP58). Several figures in this paper were created using the Python packages \texttt{numpy} and \texttt{matplotlib}.\cite{numpy}\cite{matplotlib}

\bibliography{report} 
\bibliographystyle{spiebib} 

\end{document}